\documentclass[prl,showpacs,twocolumn,amsmath]{revtex4}

\begin{document}

\title{Entanglement of Formation of Bipartite Quantum States}
\author{Kai Chen$^{1}$}
\author{Sergio Albeverio$^{1}$}
\author{Shao-Ming Fei$^{1,2,3}$}
\affiliation{$^1$Institut f\"ur Angewandte Mathematik, Universit\"at Bonn,
D-53115, Germany\\
$^2$Department of Mathematics, Capital Normal University, Beijing 100037,
China\\
$^3$Max Planck Institute for Mathematics in the Sciences,
D-04103 Leipzig, Germany}

\begin{abstract}
We give an explicit tight lower bound for the entanglement of
formation for arbitrary bipartite mixed states by using the
convex hull construction of a certain function. This is achieved
by revealing a novel connection among the entanglement of
formation, the well-known Peres-Horodecki and realignment
criteria. The bound gives a quite simple and efficiently
computable way to evaluate quantitatively the degree of
entanglement for any bipartite quantum state.
\end{abstract}

\maketitle

Quantum entangled states are used as key {\em resources} in
quantum information processing and communication, such as in
quantum cryptography, quantum teleportation, dense coding, error
correction and quantum computation \cite{nielsen}. A fundamental
problem in quantum information theory is how to
quantify the degree of entanglement in a practical and operational way
\cite{BDSW,Horo-Bruss-Plenioreviews}. One of the most meaningful and
physically motivated measures is the entanglement of formation (EOF)
\cite{BDSW,Horo-Bruss-Plenioreviews}, which quantifies the minimal
cost needed to prepare a certain quantum state in terms of EPR
pairs. Related to the EOF the behavior of entanglement has
recently been shown to play important roles in quantum phase
transition for various interacting quantum many-body systems
\cite{Osterloh02-Wu04} and may significantly affect macroscopic
properties of solids \cite{Ghosh2003}. Moreover, it has been
shown that there is a remarkable connection between entanglement
and the capacity of quantum channels \cite{Shor-Pomeransky}. A
quantitative evaluation of EOF is thus of great significance both
theoretically and experimentally.

Considerable efforts have been spent on deriving EOF or its lower bound
through analytical and numerical approaches, for some limited sets of
mixed states
\cite{Hill-Wootters97,Wootters98,Terhal-Voll2000,Vollbrecht-Werner01,%
Audenaert2001-Gittings2003,Vidal-Dur-CiracPRL2002,chenp02-Gerjuoy03-Lozinski03,%
Giedke2003,Fei-Jost2003,Fei-Jost2004,Manne-Caves2005}. Among them, the most noteworthy
results are an elegant analytical formula
for two qubits \cite{Hill-Wootters97,Wootters98}, explicit derivations for
isotropic states \cite{Terhal-Voll2000}, Werner states \cite%
{Vollbrecht-Werner01} and Gaussian states with certain
symmetries \cite{Giedke2003}. Closed-form expressions have also
been given for special classes of high
dimensional states \cite{Vidal-Dur-CiracPRL2002,Fei-Jost2003,Fei-Jost2004} and
rotationally symmetric states in $2\otimes n$ systems \cite{Manne-Caves2005},
where $n$ is the dimension of the Hilbert space associated with the second
subsystem. Possible lower bounds have been given in \cite%
{chenp02-Gerjuoy03-Lozinski03} only for $2\otimes n$ states.
Notable progress has been achieved in \cite{Mintert04,Chen-Albeverio-Fei-PRL200504}
in giving analytic lower bounds [that can be optimized further
numerically \cite{Mintert04}] for the concurrence, which permits to furnish
a lower bound of EOF for a generic mixed state. However, this lower bound
is not explicit except for the case of $2\otimes n$ systems
\cite{Chen-Albeverio-Fei-PRL200504}. For low dimensional systems, numerical methods
\cite{Audenaert2001-Gittings2003} can be used to estimate EOF. Nevertheless,
they are generally time-consuming and often not very efficient. The
notorious difficulty of evaluation for EOF is due to
the complexity to solve a high dimensional optimization problem, which becomes a
formidable task, as the dimensionality of the Hilbert space grows.

In this Letter, we present the first \emph{analytical}
calculation of a tight lower bound of EOF for \emph{arbitrary}
bipartite quantum states. An explicit expression for the bound
is obtained from the convex hull of a simple function, based on a
known result in \cite{Terhal-Voll2000}. This is achieved by
establishing a key connection between EOF and two strong
separability criteria, the Peres-Horodecki criterion \cite%
{Peres96,HorodeckiPLA96} and the realignment criterion \cite%
{Rudolph02,ChenQIC03}. The bound is shown to be exact
for some special states such as isotropic states
\cite{reductioncriterion,Terhal-Voll2000} and permits to provide
EOF estimations for many bound entangled states (BES). It
provides a very simple computable way for getting information on
the actual value of EOF, and in particular, fills significantly
the large gap between the nice result on the two qubits case
\cite{Hill-Wootters97,Wootters98} and a few other existing
results (mentioned above) for high dimensional mixed states.

Let us first recall some useful notations. A pure $m\otimes n$ $(m\leq n)$
quantum state ${\left\vert \psi \right\rangle }$ is a normalized vector in
the tensor product $\mathcal{H}_{A}\otimes \mathcal{H}_{B}$ of two Hilbert
spaces $\mathcal{H}_{A},\mathcal{H}_{B}$ for systems $A,B$. The entanglement
of formation is defined to be ${E(\left\vert \psi \right\rangle )=S(\rho
_{A})}$ where $\rho _{A}\equiv Tr_{B}({\left\vert \psi \right\rangle }%
\left\langle \psi \right\vert )$ is the reduced density matrix. Here ${%
S(\rho _{A})}$ is the entropy
\begin{equation}
{S(\rho _{A})}\equiv -\sum_{i=1}^{m}\mu _{i}\log _{2}\mu _{i}=H(\vec{\mu}),
\label{entropy}
\end{equation}
where $\mu _{i}$ are the eigenvalues of ${\rho _{A}}$ and $\vec{\mu}$
is the Schmidt vector $(\mu _{1},\mu _{2},\ldots ,\mu _{m})$. It is evident
that $E({\left\vert \psi \right\rangle })$ vanishes only for product states.
This definition can be extended to mixed states $\rho $ by the convex roof,
\begin{equation}
E(\rho )\equiv \min_{\{p_{i},|\psi _{i}\rangle \}}\sum_{i}p_{i}E({\left\vert
\psi _{i}\right\rangle }),  \label{EOF}
\end{equation}
for all possible ensemble realizations $\rho =\sum_{i}p_{i}|\psi _{i}\rangle
\langle \psi _{i}|$, where $p_{i}\geq 0$ and $\sum_{i}p_{i}=1$.
Consequently, a state $\rho $ is \emph{separable} if and only if $E(\rho )=0$
and hence can be represented as a convex combination of product states as $%
\rho =\sum_{i}p_{i}\rho _{i}^{A}\otimes \rho _{i}^{B}$, where $\rho _{i}^{A}$
and $\rho _{i}^{B}$ are pure state density matrices associated to the subsystems $A$
and $B$, respectively \cite{werner89}. The measure Eq.~(\ref{EOF}) satisfies
all the essential requirements of a good entanglement measure: convexity, no
increase under local quantum operations and classical communications on
average, no increase under local measurements, asymptotic continuity and
other properties \cite{BDSW,Horo-Bruss-Plenioreviews}.

A central idea for our approach is as follows: instead of the conventional
method to make optimization subject to a large number of constraints in Eq.~(%
\ref{EOF}), we look for minimal admissible $E({\left\vert \psi
_{i}\right\rangle })$ for a given ${\left\vert \psi _{i}\right\rangle }$
with some restrictions generated from the simple computable
Peres-Horodecki criterion of positivity under partial transpose (PPT
criterion) \cite{Peres96,HorodeckiPLA96} and the realignment criterion \cite%
{Rudolph02,ChenQIC03}. One then expects to obtain a tight lower bound
through a convex roof construction for the pure states. Without loss of
generality, we suppose that one has a general pure $m\otimes n$ quantum
state, which can always be written in the standard Schmidt form
\begin{equation}
{\left\vert \psi \right\rangle }=\sum_{i}\sqrt{\mu _{i}}{\left\vert
a_{i}b_{i}\right\rangle },  \label{Schmidt}
\end{equation}
where $\sqrt{\mu _{i}}$ $(i=1,\ldots m)$ are the Schmidt coefficients, ${%
\left\vert a_{i}\right\rangle }$ and ${\left\vert b_{i}\right\rangle }$ are
the orthonormal basis in $\mathcal{H}_{A}$ and $\mathcal{H}_{B}$, respectively.
It can be straightforwardly verified that the reduced density matrices $\rho
_{A}$ and $\rho _{B}$ have the same eigenvalues $\mu _{i}$. It follows from
Eq.~(\ref{entropy}) that $E({\left\vert \psi \right\rangle })$ vanishes only
for a product state and reaches its maximum $\log _{2}m$ for a maximally
entangled state.

Let us recollect some details on the two above mentioned criteria.
Peres made firstly a distinguished progress in the study of
separability \cite{Peres96} by showing that $\rho ^{T_{A}}\geq 0$ should be
satisfied for a separable state, where $\rho ^{T_{A}}$ stands for a partial
transpose of $\rho$ with respect to the subsystem $A$. It was further shown by
Horodecki \textit{et al.} \cite{HorodeckiPLA96} that $\rho ^{T_{A}}\geq 0$
is also sufficient for separability of $2\otimes 2$ and $2\otimes 3$
bipartite systems. In addition $\Vert \rho ^{T_{A}}\Vert $ was shown
in Refs.~\cite{Peres96,Vidal-Werner2002} to be invariant under
local unitary transformation (LU), where $||\cdot
|| $ stands for the trace norm defined by $\Vert G\Vert =Tr(GG^{\dagger
})^{1/2} $. It is clear that $\rho =\sum_{ijkl}\rho _{ik,jl}{\left\vert
a_{i}b_{k}\right\rangle }\left\langle a_{j}b_{l}\right\vert $ and $\rho
_{ik,jl}^{T_{A}}=\rho _{jk,il}$ in a suitably chosen orthonormal basis $%
a_{i} $ $(i\leq m)$ and $b_{k}$ $(k\leq n)$. Here the subscripts $i$ and $j$
can be regarded as the row and column indices for the subsystem $A$
respectively, while $k$ and $l$ are such indices for the subsystem $B$.

The \emph{realignment} criterion is another powerful operational criterion
for separability given in \cite{Rudolph02,ChenQIC03}. It demonstrates a
remarkable ability to detect many BES \cite{Rudolph02,ChenQIC03} and even
genuinely tripartite entanglement \cite{Horodecki02}. Recently considerable
efforts have been made in proposing stronger variants and multipartite
generalizations for this criterion \cite{realignmentcriteria}. The criterion
says that a realigned version $\mathcal{R}(\rho )$ of $\rho $ should satisfy
$||\mathcal{R}(\rho )||\leq 1$ for any separable state $\rho $. $\mathcal{R}%
(\rho )$ is simply defined to be $\mathcal{R}(\rho )_{ij,kl}=\rho _{ik,jl}$
\cite{Rudolph02,ChenQIC03,Horodecki02}. The LU invariant property holds also
for $||\mathcal{R}(\rho )||$ \cite{ChenQIC03}. For the pure state of Eq.~(%
\ref{Schmidt}), it is straightforward to prove
\begin{equation}
\Vert \rho ^{T_{A}}\Vert =\Vert \mathcal{R}(\rho )\Vert =\Big(\sum_{i}\sqrt{%
\mu _{i}}\Big)^{2}=\lambda ,  \label{keyobserv}
\end{equation}
as already shown in \cite{Vidal-Werner2002,Rudolph02} and \cite%
{Chen-Albeverio-Fei-PRL200504}, where $\lambda $ varies from $1$ to $m$.

Let us also review some important results in \cite{Terhal-Voll2000} which we
will use for the proof of our main Theorem. Terhal and Vollbrecht gave the
first formula for the entanglement of formation for a class of mixed states
in arbitrary dimension $d$: the isotropic states
\begin{equation}
\rho _{F}={\frac{{1-F}}{d^{2}-1}}\left( I-|\Psi ^{+}\rangle \langle \Psi
^{+}|\right) +F|\Psi ^{+}\rangle \langle \Psi ^{+}|,  \label{isotropic}
\end{equation}
where $|\Psi ^{+}\rangle \equiv \sqrt{1/d}\sum_{i=1}^{d}|ii\rangle $ and $%
F=\langle \Psi ^{+}|\rho _{F}|\Psi ^{+}\rangle $, satisfying $0\leq F\leq 1$,
is the fidelity of $\rho _{F}$ and $|\Psi ^{+}\rangle $.
They found that for $F\geq 1/d$, the EOF for isotropic
states is $E(\rho _{F})=\text{co}[R(F)]$ where $R(F)$ is a simple function
of $F$. Here the symbol \textquotedblleft co" means the convex hull, which
is the largest convex function that is bounded above by a given function.
They have also presented an explicit expression of $\text{co}[R(F)]$ for
$d=2,3 $, and conjectured its general form for arbitrary $d$.

With the above analysis and preparation, we can formulate the main result of
this Letter:

\vspace{0.1cm} \noindent {\em Theorem.---} For any $m\otimes n$ $%
(m\leq n)$ mixed quantum state $\rho $, the entanglement of formation $%
E(\rho )$ satisfies
\begin{eqnarray}
E(\rho )\geq \text{co}[R(\Lambda )],  \label{mainresult}
\end{eqnarray}
where
\begin{equation}
\left.
\begin{array}{l}
R(\Lambda )=H_{2}[\gamma (\Lambda )]+[1-\gamma (\Lambda )]%
\log _{2}(m-1), \\[2mm]
\gamma (\Lambda )=\frac{1}{m^{2}}[\sqrt{\Lambda }+\sqrt{(m-1)(m-\Lambda )}%
]^{2},%
\end{array}%
\right.  \label{Rform}
\end{equation}%
with $\Lambda =\max (\Vert \rho ^{T_{A}}\Vert ,\Vert \mathcal{R}(\rho )\Vert
)$ and $H_{2}(.)$ is the standard binary entropy function.

\vspace{0.1cm}
\emph{Proof.---}
To obtain the desired lower bound, we assume that one has already
found an optimal decomposition $\sum_{i}p_{i}\rho ^{i}$ for $\rho
$ to achieve the infimum of $E(\rho )$, where $\rho ^{i}$ are
pure state density matrices. Then $E(\rho )=\sum_{i}p_{i}E(\rho
^{i})$ by definition. For a pure state density matrix $\sigma=|\psi
\rangle \langle \psi |$ with $|\psi \rangle$ given by
Eq.~(\ref{Schmidt}), one
has $\Vert \sigma^{T_{A}}\Vert =\Vert %
\mathcal{R}(\sigma )\Vert =(\sum_{k=1}^{m}\sqrt{\mu _{k}})^{2}=$
$\lambda $ according to
Eq.~(\ref{keyobserv}). We would like firstly to find a minimal admissible $H(%
\vec{\mu})$ for a given $\lambda $. This minimization problem has
been solved in \cite{Terhal-Voll2000},
\begin{eqnarray}
R(\lambda ) &=&\underset{\vec{\mu}}{\min }\Big\{H(\vec{\mu})\mid \Big(\sum_{k=1}^{m}%
\sqrt{\mu _{k}}\Big)^{2}=\lambda \Big\}  \notag \\
&=&H_{2}[\gamma (\lambda )]+[1-\gamma (\lambda )]\log_{2}(m-1),  \label{minentropy}
\end{eqnarray}%
where
\begin{equation}
\gamma (\lambda )=\frac{1}{m^{2}}[\sqrt{\lambda }+\sqrt{(m-1)(m-\lambda )}]^{2}.
\end{equation}%
The function $R(\lambda )$ here is the $R(F)$ used in \cite{Terhal-Voll2000}
after substitutions of $d\rightarrow m$, $F\rightarrow \lambda /m$. It is
further shown in \cite{Terhal-Voll2000} that $\text{co}[R(\lambda )]$ is a
monotonously increasing, convex function and satisfies $\text{co}[R(\lambda )%
]\leq H(\vec{\mu})$ for a given $\lambda $. Denote $\mathcal{E}%
(\lambda )=\text{co}[R(\lambda )]$, one thus has%
\begin{eqnarray}
E(\rho ) &=&\sum_{i}p_{i}E(\rho ^{i})=\sum_{i}p_{i}H(\vec{\mu}^{i})%
\geq \sum_{i}p_{i}\mathcal{E}(\lambda ^{i})  \notag \\
&\geq & \mathcal{E}\Big(%
\sum_{i}p_{i}\lambda ^{i}\Big)\geq \left\{
\begin{array}{l}
\mathcal{E}(\Vert \rho ^{T_{A}}\Vert ), \\[2mm]
\mathcal{E}(\Vert \mathcal{R}(\rho )\Vert ),%
\end{array}%
\right.   \label{mainproof}
\end{eqnarray}%
where we have used the monotonicity and convexity properties of $\mathcal{E}$%
, and convexity of the trace norm $\Vert \rho ^{T_{A}}\Vert \leq
\sum_{i}p_{i}\Vert (\rho ^{i})^{T_{A}}\Vert $ and $\Vert \mathcal{R}(\rho
)\Vert \leq \sum_{i}p_{i}\Vert \mathcal{R}(\rho ^{i})\Vert $. Set $%
\Lambda =\max [\Vert \rho ^{T_{A}}\Vert ,\Vert \mathcal{R}(\rho )\Vert ]$,
we arrive at%
\begin{equation}
E(\rho )\geq \mathcal{E}(\Lambda )=\text{co}[R(\Lambda )].  \label{bound}
\end{equation}
which gives exactly the conclusion of Eq.~(\ref{mainresult}). \hfill \rule{1ex}{1ex}

Since $\text{co}[R(\lambda )]$ is the largest convex function that is
nowhere larger than $R(\lambda )$, it is optimal to give the best lower
bound according to the relations Eqs.~(\ref{minentropy}), (\ref{mainproof}%
), and (\ref{bound}). From the general form for $\text{co}[R(\lambda )]$
given in \cite{Terhal-Voll2000}, the following relation%
\begin{equation}
\begin{array}{l}
E(\rho )\geq  \\
\left\{
\begin{array}{ll}
0, & \Lambda =1, \\[2mm]
H_{2}[\gamma (\Lambda )]+[1-\gamma (\Lambda )]\log _{2}(m-1),
& \Lambda \in \big[1,\frac{4(m-1)}{m}\big], \\[2mm]
\frac{\log _{2}(m-1)}{m-2}(\Lambda -m)+\log _{2}m, & \Lambda \in \big[\frac{%
4(m-1)}{m},m\big],
\end{array}%
\right.
\end{array}
\label{lowerbound}
\end{equation}
holds for $m=2,3$. We have strong evidence for its correctness for arbitrary
$m$ by verifying directly, that the second derivative for
$R(\lambda )$ with respect to $\lambda $ goes from positive to
negative value with only one zero point when $\lambda $ varies
from $1$ to $m$ according to the analysis of
\cite{Terhal-Voll2000}.

The result of the above Theorem and its general expression Eq.~(\ref{lowerbound})
provide an explicit tight lower bound for the EOF without the need
of any numerical optimization procedure. In fact, it can be done in an entirely
straightforward manner through the computation of the trace norm of a certain
matrix by standard linear algebra packages. Some further
significant features arising from our general result are
illustrated in several examples and the following discussions.

\vspace{0.1cm} \emph{Example 1:} Qubit-qudit system

When $m=2$, which corresponds to a qubit-qudit system, one derives easily
from the Theorem and Eq.~(\ref{lowerbound}) that
\begin{equation}
E(\rho )\geq \left\{
\begin{tabular}{ll}
$0$, & $\Lambda =1$, \\
$H_{2}[\frac{1}{2}\big(1+\sqrt{1-(\Lambda -1)^{2}}\big)]$, & $%
\Lambda \in \big[1,2\big]$.%
\end{tabular}
\right.  \label{2n}
\end{equation}
This recovers previous results obtained by the authors in \cite%
{Chen-Albeverio-Fei-PRL200504}, where $\Lambda =\max [\Vert \rho
^{T_{A}}\Vert ,\Vert R(\rho )\Vert ]$, and others in \cite{Chi-Lee2003}, where
$\Lambda =\Vert \rho ^{T_{A}}\Vert $. In addition Eq.~(\ref{2n}) can detect and give
lower bounds of EOF for \emph{all} entangled states of two qubits
and of qubit-qutrit systems, since the Peres-Horodecki criterion is necessary
and sufficient for separability in these cases \cite{HorodeckiPLA96}.
Furthermore, whenever there is a two qubit state in which $\Lambda -1$ is
equal to the concurrence defined in \cite{Wootters98}, the bound Eq.~(\ref%
{2n}) will give the exact value of EOF. For example the $2\otimes 2$ Werner state
\cite{Vollbrecht-Werner01} fits this condition by direct verification.
The bound Eq.~(\ref{2n}) will be particularly useful for the
study of entanglement in realistic many-body physical systems. For
example, one usually needs to monitor entanglement dynamics and distribution
between a spin $1/2$ particle and the remaining parts for a solid state
system or a quantum computing device, and our bound can be useful in this context.

\vspace{0.1cm}
\emph{Example 2:} Isotropic states

Isotropic states Eq.~(\ref{isotropic}) were firstly proposed
in \cite{reductioncriterion} and further
properties were analyzed in \cite{Vollbrecht-Werner01}. They arise naturally
in some special depolarizing channel \cite{reductioncriterion} and constitute the class of
$U\otimes U^{\ast }$ invariant mixed states in $d\otimes d$ systems.
These states have been shown to be separable for $F\leq 1/d$ \cite{reductioncriterion}.
The EOF $E(\rho )$ for this class of states has been given in \cite%
{Terhal-Voll2000} by an elegant extremization procedure. It is derived in
\cite{Vidal-Werner2002,Rudolph02} that $\Vert \rho _{F}^{T_{A}}\Vert =\Vert
\mathcal{R}(\rho _{F})\Vert =dF$ for $F>1/d$. Thus one can directly exploit
the above Theorem with $\Lambda =dF$ to see that the bound given
in Eqs.~(\ref{mainresult})
and (\ref{lowerbound}) coincides with the exact value of EOF for the whole
class of states in Eq.~(\ref{isotropic}).

\vspace{0.1cm} Our Theorem and the general relation of Eq.~(\ref{lowerbound})
complement a number of existing approaches to
make a quite good estimate of entanglement for BES, benefiting from the
powerful realignment criterion which enables one to detect many of the BES \cite%
{Rudolph02,ChenQIC03}.

The bound can be made even better if it is much easier to compute a
convex-roof extended entanglement measure \cite{Rudolph02,Soojoon2003} $%
\Vert \rho ^{T_{A}}\Vert _\text{co}$ or $\Vert \mathcal{R}(\rho )\Vert _\text{co}$,
than the EOF. The extended measures in our case are defined by $\Vert \rho
^{T_{A}}\Vert _\text{co}\equiv \min_{\{p_{i},\rho ^{i}\}}\sum_{i}p_{i}\Vert (\rho
^{i})^{T_{A}}\Vert $ and $\Vert \mathcal{R}(\rho )\Vert _\text{co}\equiv
\min_{\{p_{i},\rho ^{i}\}}\sum_{i}p_{i}\Vert \mathcal{R}(\rho ^{i})\Vert $
where $\rho =\sum_{i}p_{i}\rho ^{i}$, $p_{i}\geq 0$ and $\sum_{i}p_{i}=1$.
They have been studied and calculated for some special class of states in
\cite{Rudolph02,Soojoon2003}. Defining $\Lambda =\max (\Vert \rho
^{T_{A}}\Vert _\text{co},\Vert \mathcal{R}(\rho )\Vert _\text{co})$, one finds that
the result of the Theorem is still valid, since the last inequality in Eq.~(%
\ref{mainproof}) holds as $\mathcal{E}\Big(\sum_{i}p_{i}\lambda ^{i}\Big)%
\geq \mathcal{E}(\Lambda )$. This will provide a tighter lower bound for
the EOF since generally $\Vert \rho ^{T_{A}}\Vert _\text{co}\geq \Vert \rho
^{T_{A}}\Vert $ and $\Vert \mathcal{R}(\rho )\Vert _\text{co}\geq \Vert \mathcal{R%
}(\rho )\Vert $ follow from the definitions.

Due to the nonanalytic behavior of the right hand side of
Eq.~(\ref{lowerbound}), it is difficult to find a specific
condition under which our bound will be exact for a general
state. Roughly, for a $2\otimes n$ system one necessary
requirement is that all the $\rho ^{i}$ should be equally
entangled with $E(\rho ^{i})=R(\Lambda )$ in an optimal
decomposition for achieving EOF, as in this case we demand all
the inequalities in Eq.~(\ref{mainproof}) to be changed into
equalities. In higher dimensions, it is necessary that all the $\rho ^{i}$
should have equal EOF of $R(\Lambda )$ when $1\leq \Lambda \leq
4(m-1)/m$ while there
should be two values of EOF, $\log _{2}m$ and $R\big(4(m-1)/m\big)$ for $%
4(m-1)/m\leq \Lambda \leq m$, as seen from Eq.~(\ref{lowerbound}).

Although the work \cite{Chen-Albeverio-Fei-PRL200504} permits to
furnish a lower bound for EOF (in fact there is no explicit
formula given there), it is by no means optimal for general
states except for $2\otimes n$ systems. Instead, the procedure in
\cite{Chen-Albeverio-Fei-PRL200504} clearly imposes more
restrictions than that of the present work: it requires firstly
to give a lower bound of concurrence for a given $\Lambda $, and
then obtain a possible lower bound of EOF from the derived
concurrence bound. Thus the result in the present Letter is
optimal as long as the parameter $\Lambda $ is involved only,
since we have utilized the largest convex function
$\text{co}[R(\lambda )]$ that is bounded above by $R(\lambda )$.

In summary, we have determined a completely analytic lower bound
of EOF for an arbitrary bipartite mixed state, which
characterizes optimally the quantitative behavior of entanglement
through the well-known Peres-Horodecki criterion and the
realignment criterion for separability. The procedure only
involves a simple computation of matrix eigenvalues and can be
done efficiently with the standard linear algebra packages. Our
bound leads to exact values of EOF for some special quantum
states and enables one to give an easy EOF evaluation for many
BES, a task which was extremely difficult before. We are of the
opinion that the result constitutes a significant bridge over the
big gap between the elegant result of Wootters for 2 qubits, and
the few existing results mentioned before for high dimensional
states. In this way our method can yield a powerful tool for
investigating quantitatively the character of entanglement for
practical laboratory sources (atomic, photonic, spin or other
carriers), and for the study of many-body systems. Furthermore,
it may provide important insights into realistic quantum channels
and condensed matter systems, revealing deep connections between
entanglement and macroscopic properties for the corresponding
physical systems.

K.C. gratefully acknowledges support from the Alexander von
Humboldt Foundation. The support of this work by the Deutsche
Forschungsgemeinschaft SFB611 and German(DFG)-Chinese(NSFC)
Exchange Programme 446CHV113/231 is also gratefully acknowledged.
Partially supported by NKBRPC (2004CB318000).

\end{document}